# Decentralized AI-driven IoT Architecture for Privacy-Preserving and Latency-Optimized Healthcare in Pandemic and Critical Care Scenarios

Completed Research Paper

**Harsha Sammangi**
Ph.D Candidate, IS and Analytics
Dakota State University
Sammangi.harsha@trojans.dsu.edu

**Giridhar Reddy Bojja, Ph.D.**
Assistant Professor of IS and Analytics
College of Business
Michigan Technological University
gbojja1@mtu.edu

Aditya Jagatha
College of Business and Information Systems
Dakota State University
aditya.jagatha@trojans.dsu.edu

Dr. Jun Liu, Ph.D
Professor of M.S and Analytics
College of Business & I.S
Dakota State University
Jun.liu@dsu.edu

## Abstract

AI Innovations in the IoT for Real-Time Patient Monitoring On one hand, the current traditional centralized healthcare architecture poses numerous issues, including data privacy, delay, and security. Here, we present an AI-enabled decentralized IoT architecture that can address such challenges during a pandemic and critical care settings. This work presents our architecture to enhance the effectiveness of the current available federated learning, blockchain, and edge computing approach, maximizing data privacy, minimizing latency, and improving other general system metrics. Experimental results demonstrate transaction latency, energy consumption, and data throughput orders of magnitude lower than competitive cloud solutions.

## Introduction

Combining of AI and IoT in healthcare is a paradigm shift in patient monitoring, diagnosis, and treatment. Wearables, remote monitoring systems and AI-driven analytics allow healthcare services to have access to patients' real-time data anywhere in the world in an instant, thereby breaking down silos, and ending isolated reports. We are still going to be using that digitization with an eye towards the future of preventive care, rapid-response emergencies, and clinical decision-making. But behind all these innovations lies a problem that is not going away: dependence on centralized health systems. These systems sufficed, but not for modern healthcare; speed, security, and scalability must be non-negotiable. In this paper, we investigate these challenges and provide a decentralized, AI-driven solution with the potential to revolutionize healthcare delivery in the coming decades.

Well-known problems abound with typical CloudIoT-based centralized healthcare systems. Another big challenge is privacy, poor data collection and storage practices result in sensitive health data being leaked (Zhang et al., 2021). Another challenge to the complexity in networking latency—it prolongs the time to transmit data fetched by distributed IoT devices to far-off locations servers which could be a catastrophe in time-constrained medical and clinical cases. Scalability is another bottleneck, Centralized databases have been buckling under the exponential growth of IoT-generated health data. Beyond technical restrictions, these systems undermine trust, cost patients and providers data ownership, and fail to prove compliance with statutes including the GDPR and HIPAA (Tanwar et al., 2020). In high-stakes moments

— pandemics, emergencies — those challenges are multiplied, in which secure, quick access to health information can literally mean the difference between life and death.

This proposed research represents a Secure and Decentralized AI-driven IoT Architecture covering the aforementioned limitations on centralized models. We propose an all-encompassing solution to this issue, which integrates a blockchain-based federated learning technology with edge computing resources and AI-based anomaly detection. Hyperledger Fabric also provides tamper-proof data integrity and transparent access control and federated learning assists in training the models on the IoT devices at their respective locations to ensure patient privacy. Edge computing addresses latency by minimizing the distance data must travel from its point of origin to insights generated in near real time,

In addition, more advanced security practices — such as differential privacy, homomorphic encryption and zero-knowledge proofs — make breaches less of a concern. Smart contracts, for instance, also serve to enforce trust, and limit access to authorized personnel only (according to some policy). This comprehensive perspective bridges technical gaps and aligns with practical needs of contemporary healthcare systems. This paper contributes theoretically and practically. We introduce a new framework that employs blockchain and AI to achieve scalable, near real-time automation of health care procedures, which we validate with a prototype that incorporates IoT sensors, federated learning, and Hyperledger Fabric. Experimental results highlight the potential: Our architecture improves latency, data throughput, and energy efficiency by 40%, 120% and 30%, respectively, compared to traditional cloud-based architectures. This research ideally caters to emergency care and privacy-sensitive scenarios and lies at the intersection of AI, blockchain, and IoT while providing an effective secure, efficient, and trust-enhancing solution. As healthcare's needs change, our work provides a forward-looking roadmap to build resilient, patient-centered systems in ways that could be extended at scale to a broader number of uses.

# Related Work

Significant attention by scholars has been placed on the integration of artificial intelligence (AI), the Internet of Things (IoT), and blockchain in health care due to the ability to protect, have low latency, and maintain privacy of patient data. Centralized, decentralized, and hybrid approaches of optimization of healthcare IoT systems architecture have been studied in several research. In this section, we discuss the related works, which consists of different architectures of healthcare IoT systems, AI and federated learning, Blockchain-based paradigms in healthcare, and privacy-preserving algorithms.

## Centralized Healthcare Internet of Things Systems and Their Security Limitations

The traditional approach to health care IoT has been to use centralized architectures (i.e., cloud based architectures) where patient records and the data collected using IoT devices were all stored on remote servers. Studies such as Zhang et al. (2021) and Gupta et al. (2022) acknowledge the benefits of scalability, but point out that it can be associated with serious downsides: security vulnerabilities, privacy risks and processing delays. However, these databases aren't without abuse and there have been massive leaks exposing hundreds of millions of records. make concurrent emergency, uncertainty and real time decision making challenge in critical care settings. Latency compounds these problems as it can take time to send data to central servers, which would render these models unusable in an emergency. (Sharma et al., 2021). These shortcomings highlight the pressing need for decentralized alternatives.

## Healthcare Analytics Utilizing AI-Based Federated Learning

Healthcare analytics have been improved by AI, enabling anomaly detection and predictive insight.(Tang et al., 2020). However, classical methodology in machine learning, characterized by centralized data aggregation, stands in opposition to privacy regulations like HIPAA or GDPR. (Wang et al., 2021). In contrast, Federated Learning (FL) provides a privacy-preserving solution, allowing distributed devices to collaboratively train models without sharing raw patient data across the system. (McMahan et al., 2017). Previous work (Kairouz et al., 2021) (Lyu et al., 2022) shows the trade-off between security and performance of FL. However, there are still many challenges, including data heterogeneity, communication overhead, and the communication threshold between in-network attacks (Yang et al., 2022). FL successes in security internals, but the trade-off for efficiency remains to be seen.

## Health Frameworks Based on Blockchain

Blockchain technology has thus far been extensively studied to address issues related to healthcare IoT, such as data integrity, trust, and decentralized access control. (Kassab et al. 2020) Moreover, (Mendhurwar et al., 2021) Utilize smart contracts to facilitate and secure transparent data sharing. However, scalability and computational overheads are considerable hurdles. Traditional consensus mechanisms like Proof of Work (PoW) employed in Bitcoin come with heavy resource demands that are inappropriate for real-time healthcare (Zhao et al., 2021). To resolve this issue, various lighter alternatives of PoW have been suggested in the literature, such as Proof of Authority (PoA) and Delegated Proof of Stake (DPoS)(Rahman et al., 2022). However, their incorporation into healthcare systems still needs to be improved.

## Privacy-Preserving Methodologies in Healthcare IoT

Privacy-Preserving Methodologies in Healthcare IoT Securing sensitive patient data continues to be one of the pillars of healthcare research. (Sun et al. 2021) Moreover, (Chen et al., 2022) Criticize traditional encryption styles and promote revolutionary cryptographic approaches such as homomorphic encryption (HE), secure multi-party computation (SMPC), and zero-knowledge proofs (ZKP) for maximum security digits. Recent efforts, such as (Basu et al., 2023), have found a way to apply differential privacy on top of federated learning, injecting noise into the training data to obfuscate where data comes from without sacrificing accuracy in the model. Nevertheless, these techniques may often lead to increased computational complexity, requiring further optimization for deployment at scale. (Liu et al., 2023). Maintaining both privacy and efficiency is still an unsolved challenge.

### Existing Shortcomings and Contributions of this Study

Critical gaps persist despite advances in blockchain, AI, and privacy-preserving techniques. First, latency optimization in decentralized architectures lags, limiting their applicability to real-time critical care (Hassan et al., 2022). Second, scalability and interoperability across healthcare providers remain underexplored, hindering widespread adoption (Patel et al., 2021). Third, trust and access control mechanisms lack automation and policy-driven precision (Jiang et al., 2022). This study addresses these shortcomings with a Secure and Decentralized AI-driven IoT Framework, integrating (1) federated learning for privacy-compliant model training, (2) blockchain with smart contracts for automation and secure access control, and (3) edge computing for low-latency, real-time anomaly detection. By bridging these gaps, our work advances the state of the art in healthcare IoT.

## Proposed Framework

### System Architecture

Here, we present a Secure and Decentralized AI-driven IoT Architecture for Health, in which security and adaptivity are top priorities based on our previous work. This paper outlines a framework that integrates the key ideas of the Internet of Things (IoT), edge computing, federated learning, blockchain, and privacy-preserving cryptographic mechanisms into a secure, scalable, and efficient mechanism for maintaining healthcare data. You are taught a four-layer architecture in which each layer takes responsibility for a set of functions for acquiring, processing, securing, and controlling access to data.

Layer 1: IoT Data Collection and Edge Processing

At the core of the architecture are IoT-enabled medical sensors, such as wearables and remote monitoring devices that record real-time patient vitals (heart rate, blood pressure, oxygen saturation, ECG, etc.). Instead of sending raw data to centralized servers over the cloud (which can incur latency and compromise the privacy and security of users' sensitive information), this framework uses edge computing. Edge nodes (e.g., NVIDIA Jetson boards, Raspberry Pi clusters, intelligent hospital gateways) preprocess data locally, filter out noise, extract essential features, and perform preliminary anomaly detection. The approach conserves bandwidth by preventing unnecessary data transmissions, avoiding network congestion, and facilitating real-time responses in urgent situations, like abnormal vitals detection, leading to prompt intervention.

Layer 2: Federated Learning and GPT Model Training

Layer 2 utilizes federated learning (FL) to replace central data aggregation prevention and the subsequent privacy risks. Through a method where distributed edge and hospital nodes can train AI models cooperatively but locally without feeding raw patient data across the wire. The workflow consists of three steps: (1) local training on patient health data from each data owner/healthcare organization, (2) transference of the model weight updates (not the raw data) with a blockchain-secured aggregator, (3) global aggregation of the model aimed at improving performance without privacy sacrifices. Differential privacy (DP) provides more robust security guarantees by adding controlled noise to updates, preventing adversaries from reconstructing input features. This layer makes it possible to have HIPAA and GDPR compliance without sacrificing advanced, privacy-preserving diagnostics.

Layer 3: Blockchain And Smart Contracts For Secure Data Management

Through Layer 3, Hyperledger Fabric permissioned blockchain technology is integrated to provide data integrity, privately secure FL model updates, and support decentralized trust. Overcoming data silos and interoperability pain points, it creates tamper-proof records of every transaction—from AI model updates or data access requests to modifications. Primary features include: (1) an immutable hashed log of all transactions, (2) policy-based rules enforced via role- and attribute-based access control executed by smart contracts, (3) pervasive-decentralized authentication achieved through self-sovereign identity, and (4) low- latency healthcare application specific Proof of Authority (PoA | consensus mechanism. In contrast to automated PoW, PoA is a lightweight mechanism with minimal computational overhead and can, therefore, easily be used in real time.

Layer 4: Access control and privacy-preserving mechanisms

This system ensures that the patient data is secured in the top layer and accessible only through advanced cryptography by whom the patient allows. In contrast to traditional techniques that risk record breaches, this framework counters with (1) Zero-Knowledge Proofs (ZKP) that confirm the genuineness of data without disclosing contents and (2) Homomorphic Encryption (HE) for permit computations on encrypted data while enabling analytics without compromising confidentiality (3) Attribute-Based Access (ABAC) that implements fine-grained, role-based policies. Exposing FHIR-compliant APIs enables architecture to facilitate interoperability across the spectrum from hospitals to research institutions and telemedicine platforms while ensuring data sovereignty and regulatory compliance.

**Integration and Benefits**

Its multi-layer design sequentially addresses healthcare IoT challenges: edge computing for low latency, FL with DP for privacy-preserving AI, blockchain for creating immutable audit trails, and smart contracts for automating secure access. Designed with pandemic response, remote monitoring, and AI-assisted diagnostics in mind, it paves the way for scalable, autonomous healthcare systems.

## Privacy and Security Mechanisms

### Threat model and the security challenges

Federated Learning is a paradigm where local models are trained on respective data sources, and training updates (not the raw data) are sent to a central server for model aggregation. DP also strengthens this by introducing noise into an update, making inference attacks and reconstructing the raw data impossible. Advantages include (1) instant HIPAA/GDPR compliance through on-site data retention, (2) resilience against AML attacks through decentralized training, and (3) protection from inference-versus-attack, maintaining patient anonymity even upon inspection.

### Data Integrity and Secure Transactions via Blockchain

Data integrity and transparency on Hyperledger Fabric without transaction immutability, distributed trust, and auditable trails prevent unauthorized data access. PoA consensus lowers energy costs compared to PoW, matching healthcare's real-time requirements.

### Access Control with Smart Contracts and Attribute-Based Authorization

Smart contracts automate policy-driven permissions on-chain, while attribute-based access control allows for dynamic role-based access control off-chain. Data access is restricted to authorized personnel, with access policies tailored to specific institutional and regulatory needs to mitigate the potential of unauthorized exposure.

Additional Cryptographic Techniques: Homomorphic Encryption and Zero-Knowledge Proofs

Allows analytics on encrypted data without decryption, and ZKP verifies transactions without record exposure. Combined, they provide processing confidentiality and authentication, increasing system privacy.

### Secure communication and management

This can be done by implementing end-to-end encryption (E2EE) using Elliptic Curve Cryptography (ECC) and Diffie-Hellman key exchange between IoT devices, edge nodes, and blockchain peers. ECC is lightweight, so it is suitable for resource-constrained devices where computational power usage without compromising security is a concern.

Threat Mitigation Approaches and Compliance with Regulation

The framework incorporates (1) HIPAA/GDPR adherence via privacy-first design, (2) ML-based intrusion detection systems (IDS) to monitor network anomalies, and (3) regular adversarial testing to harden AI models against poisoning. These measures ensure regulatory compliance and system resilience.

| Security Feature | Purpose | Technology Used |
|---|---|---|
| Federated Learning | Decentralized AI model training | Localized AI model updates |
| Differential Privacy | Privacy-preserving AI training | Noise injection into model weights |
| Blockchain | Data integrity and transparency | Hyperledger Fabric |
| Smart Contracts | Automated access control | Ethereum-based execution |
| Attribute-Based Access Control (ABAC) | Policy-driven authorization | Dynamic role-based policies |
| Homomorphic Encryption (HE) | Privacy-preserving computations | Encrypted AI analytics |
| Zero-Knowledge Proofs (ZKP) | Secure authentication | Cryptographic proofing |
| Elliptic Curve Cryptography (ECC) | Secure IoT data transmissions | Lightweight cryptographic keys |

## Workflow

The framework operates as an end-to-end pipeline:

- Data Collection and Preprocessing: IoT sensors gather vitals and edge nodes, preprocess and anonymize data using DP and ECC encryption.

- Federated Learning: Local nodes train AI models, share DP-protected updates via Hyperledger Fabric, and aggregate globally.

- Blockchain Transactions: Smart contracts log and immutably secure all actions—model updates and access requests.

- Access Control: ABAC and ZKP authenticate users, and HE enables encrypted analytics.

- Real-Time Decision-Making: AI detects anomalies, smart contracts trigger alerts, and authorized providers access records for swift intervention.

This workflow ensures privacy-preserving, low-latency healthcare delivery, validated in subsequent experimental analysis.

## Experimental Analysis

### Testbed Configuration

To assess the efficacy of the proposed Secure and Decentralized AI-driven IoT Framework, we established a real-world testbed integrating IoT-enabled medical sensors, edge computing infrastructure, federated learning servers, and a Hyperledger Fabric blockchain. This setup was designed to emulate a healthcare environment where patient data is continuously captured, processed at the edge, securely shared via federated learning, and immutably recorded on a blockchain ledger. The primary goals were to evaluate the framework's latency, scalability, privacy preservation, and computational efficiency under diverse operational conditions.

The testbed included an Internet of Things (IoT)-based health monitoring system that contained medical sensors (e.g., ECG wearable monitors, pulse oximeters, blood pressure monitors, glucose sensors) for real-time encrypted transmission of patient vitals to edge computing nodes. The lightweight AI inference, anomaly detection, and differential privacy transformations were performed in external nodes based on NVIDIA Jetson Nano and Raspberry Pi clusters. Then, model updates from the edge were sent to a federated learning server.

We developed a federated learning architecture with TensorFlow Federated (TFF), allowing distributed healthcare nodes to train AI in a privacy-preserving manner. By drawing on decentralized hospital databases, the system compiled insights — for example, the efficacy of COVID-19 treatments across geographies — without moving raw patient data and using Hyperledger Fabric blockchain for synchronization through secure storing of federated learning updates, verifying patient records, and enforcing decentralized access control. A Hyperledger Fabric peer was deployed at each hospital node to serve federated learning and AI-powered anomaly detection.

Smart contracts written in Solidity were deployed on the blockchain to enforce policy-driven access control, automate consent management, and log medical transactions immutably. Furthermore, an Ethereum-based Zero-Knowledge Proof (ZKP) authentication layer enabled only authorized healthcare providers to decrypt and read patient records, thus protecting sensitive information. The testbed was deployed in a hybrid cloud environment across Microsoft Azure and Amazon AWS, simulating geographically distributed healthcare facilities. At a central node, local models encrypted at distributed cloud nodes were aggregated without exchanging raw data.

The performance of the framework was evaluated compared to traditional cloud-based healthcare architectures, taking into account transaction latency, energy consumption, computational overhead, and model accuracy. For that, three healthcare use cases were tested:

- Emergency Critical Care Monitoring: Edge nodes monitored patient vitals in real time (e.g. abnormal ECG readings or sudden blood pressure drops), sending situation-awareness alerts through smart contracts.

- Remote Telemedicine Consultations: Blockchain authentication secured privacy-preserving AI diagnostics access for remote physicians.

- Predicting Chronic Disease Using Federated Learning Across Hospitals: Metrics for diabetes, cardiovascular and respiratory diseases were predicted by storing the data in each hospital with an emphasis on early intervention.

This decentralized, privacy-preserving, real-time healthcare architecture achieved lower latency, improved security, and scalable AI-enabled analytics. Next, we describe its behavior under different parameters.

**Performance Evaluation**

To assess the framework's performance to be suitable for real-time healthcare applications, AI-based diagnosis, and distributed medical data management, we evaluated the framework in the four main metrics—Transaction latency, throughput, privacy efficiency, and energy consumption under various operating conditions. Transaction Latency Analysis

Latency is a sensitive subject in healthcare IoT systems as it is fundamental in emergency cases that require real-time responses. High latency is a consistent issue in traditional cloud architectures due to centralized processing and longer data transfers. By contrast, our framework avoids delays through edge computing and blockchain smart contracts. Experimental results demonstrated the reduction of transaction latency by 40% when compared to the traditional cloud models; for example, in emergency medical response scenarios, smart contracts yielded average latency times 120 milliseconds, compared with greater than 200 milliseconds in central systems in which delays are often amplified by network congestion. These were facilitated through localized edge AI processing, decentralized decision-making with federated learning, and automated blockchain access control.

**Throughput and Scalability**

By observing the throughput (the number of transactions processed successfully per second (TPS)), we can gauge the system's scalability. Older blockchain networks, like Ethereum, with Proof of Work (PoW) consensus, show low throughput. Our framework exhibited 120 against the signature of an individual patient on Hyperledger Fabric with Proof of Authority (PoA) consensus, which was a 150% improvement over the conventional cloud-based electronic health record (EHR) systems. The PoA mechanism speeds up block confirmation, allowing medical clicks and AI model updates to be recorded in real time without a computing bottleneck. Performance and security assessment

Data privacy and security are significant challenges in healthcare IoT. We provide a new paradigm for applying differential privacy with federated learning and ZKPs for distributed auditable data processing by integrating them into a flexible framework. The resistance of the system data reconstruction attacks and compliance with regulations such as HIPAA and GDPR evaluated efficiency in privacy. In experiments, the differential privacy-enhanced federated learning blocked 92% of adversarial inference attacks that attempted patient data reconstruction from model updates. Compared to traditional systems like role-based access control (RBAC), the keyless password (ZKP) based authentication reduced unauthorized access by over 95%. We then apply the approval process in a permissioned Hyperledger Fabric blockchain with homomorphic encryption to prevent insider and clever contract attacks, guarantee data integrity, and perform AI computation on this encrypted data.

**Optimized Energy Consumption**

Energy efficiency is vital for IoT-based healthcare systems, particularly for wearable devices and remote monitoring. Cloud-based AI architectures often incur high energy costs. Our framework, utilizing edge-based AI inference and PoA consensus, reduced energy consumption by 30% compared to centralized cloud models. This reduction resulted from localized AI inference minimizing cloud dependency, federated learning eliminating centralized training, and PoA replacing energy-intensive PoW mechanisms. Additionally, optimized energy use in IoT sensors and edge devices extended the operational life of battery-powered wearables. Comparative Study With Legacy Systems

The applicability and potential of the framework were assessed through a comparative exploration with existing cloud-based and blockchain-based healthcare architectures. The results are below in the table:

| Metric | Proposed Framework | Traditional Cloud-Based AI | Standard Blockchain (PoW) |
|---|---|---|---|
| Transaction Latency | 120ms | 200ms | 600ms |
| Throughput (TPS) | 120 TPS | 50 TPS | 15 TPS |
| Data Privacy Protection | High (Federated Learning & ZKP) | Low | Moderate |
| Energy Consumption | Low (30% lower) | High | Very High |
| Security Against Attacks | Highly Secure (HE, DP, ZKP) | Moderate | High |

# Conclusion

This study introduces a Secure and Decentralized AI-driven IoT Framework to address healthcare systems' privacy, security, latency, and interoperability challenges. The framework ensures scalable, secure, and efficient healthcare data management by integrating blockchain, federated learning, and edge computing. It reduces latency by 40% (120 TPS), cuts energy use by 30%, and enhances privacy-preserving AI analytics while securing data access with blockchain-based controls like Zero-Knowledge Proofs (ZKP) and homomorphic encryption (HE). Patient data remains local, contributing to global AI models without compromising regulatory compliance (e.g., HIPAA, GDPR). Experimental results highlight its effectiveness for real-time applications, including emergency response, telemedicine, and predictive analytics. However, challenges in distributed learning and the need for post-quantum cryptography remain, alongside opportunities to leverage 6G and energy-efficient AI for broader adoption.

## Future Research

Future efforts should enhance federated learning security with adversarial-resistant models, incorporating secure aggregation, differential privacy, and anomaly detection to counter data poisoning and model threats. Developing post-quantum cryptography—such as lattice-based methods and quantum-safe protocols, will strengthen blockchain resilience against emerging quantum risks. Research should focus on lightweight AI models for edge inference and low-energy consensus mechanisms like Proof of Stake (PoS) or Directed Acyclic Graphs (DAGs) to boost energy efficiency. With AIoT on the horizon, integrating intelligent networking, dynamic resource allocation, and adaptive security will improve scalability, latency, and interoperability in decentralized healthcare systems.